\documentclass{article}

\usepackage{arxiv}

\usepackage[utf8]{inputenc} % allow utf-8 input
\usepackage[T1]{fontenc}    % use 8-bit T1 fonts
\usepackage{hyperref}       % hyperlinks
\usepackage{url}            % simple URL typesetting
\usepackage{booktabs}       % professional-quality tables
\usepackage{amsfonts}       % blackboard math symbols
\usepackage{nicefrac}       % compact symbols for 1/2, etc.
\usepackage{microtype}      % microtypography
\usepackage{lipsum}

\usepackage{amsmath, setspace,geometry,color}
\usepackage{amssymb}
\usepackage{graphicx}
\usepackage{xcolor}
\usepackage[ruled, boxed, vlined]{algorithm2e}
\usepackage{algorithmic}
\usepackage{multirow}
\usepackage[caption=false,font=footnotesize]{subfig}
\usepackage{url}

\newtheorem{definition}{Definition}

\title{Novel Radiomic Feature for Survival Prediction of Lung Cancer Patients using Low-Dose CBCT Images}

\author{
  Bijju Kranthi Veduruparthi\textsuperscript{*}, Jayanta Mukherjee, Partha Pratim Das\\
  Department of Computer Science and Engineering\\
  Indian Institute of Technology Kharagpur, India\\
  \textsuperscript{*}\texttt{bijjuair@gmail.com} \\
   \And
   Moses Arunsingh, Sanjoy Chatterjee  \\
   Department of Radiation Oncology\\
   Tata Medical Center, Kolkata, India\\
   \And
   Raj Kumar Shrimali \\
   Arden Cancer Centre\\
   University Hospitals Coventry and Warwickshire\\
   NHS Trust, Coventry, United Kingdom\\
   \And
   Sriram Prasath\\
   Department of Medical Physics\\
   Tata Medical Center, Kolkata, India\\
   \And
   Soumendranath Ray\\
   Department of Nuclear Medicine\\
   Tata Medical Center, Kolkata, India\\
}

\begin{document}
\maketitle

\begin{abstract}
Prediction of survivability in a patient for tumor progression is useful to estimate the effectiveness of a treatment protocol.
In our work, we present a model to take into account the heterogeneous nature of a tumor to predict survival.
The tumor heterogeneity is measured in terms of its mass by combining information regarding the radiodensity obtained in images with the gross tumor volume (GTV).
We propose a novel feature called Tumor Mass within a GTV (TMG), that improves the prediction of survivability, compared to existing models which use GTV.
Weekly variation in TMG of a patient is computed from the image data and also estimated from a cell survivability model.
The parameters obtained from the cell survivability model are indicatives of changes in TMG over the treatment period.
We use these parameters along with other patient metadata to perform survival analysis and regression.
Cox's Proportional Hazard survival regression was performed using these data.
Significant improvement in the average concordance index from 0.47 to 0.64 was observed when TMG is used in the model instead of GTV.
The experiments show that there is a difference in the treatment response in responsive and non-responsive patients and that the proposed method can be used to predict patient survivability. 
\end{abstract}

% Include a list of up to six keywords after the abstract
\keywords{Cone Beam Computed Tomography \and Jacobian \and Gross Tumor Volume \and Radiomics \and CT Number \and Image Registration \and Survival Regression \and Cox's Proportional Hazard model}

\section{Introduction}
\label{sect:intro}  % \label{} allows reference to this section
Progression of a tumor is indicated in an image in terms of change in tumor size, shape, and extent of spread to different regions of the anatomy.
A less aggressive tumor is bound to a particular region, while a more aggressive tumor is likely to spread to several organs denoting metastasis.
Aggressive tumors adversely affect patient health leading to lower survival time.
Therefore, the rate of tumor volume was observed to correlate to survival in several works.
Based on several observations of the tumor volume, a period of survival with or without progression can be calculated.

In this work, the problem of prognosis is framed as a regression problem.
The regression variable is defined in terms of the Progression Free Survival (PFS).

\begin{definition}
The shortest time between treatment completion and progression of tumor in any subsequent follow up is called the Progression Free Survival (PFS).
\end{definition}

It is a measure of the effectiveness of the treatment protocol in treating the patient.
A low PFS indicates, that the treatment protocol could have been different.
Early detection of PFS, just after the completion of treatment, can help the doctors take alternative steps to cure the patient, or to treat the patient in such a way that the PFS increases.
Other measures, like the Overall Survival (OS), are also used to see how long the patient has actually survived due to the treatment.
In this work, PFS is used to measure the survivability of a patient.

\subsection{Related Works}
Survival analysis involves making predictions about the survival time and estimation of survival probability at the estimated survival time.
The approaches of survival analysis can be categorized into statistical, and machine learning based methods \cite{collett2015modelling, Wang2017, parmar2015machine}.
Statistical methods focus on regressing survival curves, while in machine learning based approaches, the target is to estimate the time of occurrence of the event, i.e. progression of a tumor.
Statistical methods for survival analysis can be parametric, non-parametric or semi-parametric \cite{miller2011survival}.
Techniques performing Kaplan-Meier estimate \cite{kaplan1958nonparametric} and Nelson-Aalen estimate \cite{nelson1969hazard} are typical examples of non-parametric methods.
Semi-Parametric methods like Cox regression \cite{cox2018analysis}, with and without time-dependency are widely used.
In machine learning based methods, random survival forests \cite{Ishwaran2008}, survival trees \cite{bou2011review,leblanc1993survival,gordon1985tree}, Bayesian methods \cite{ibrahim2001bayesian}, neural networks \cite{biganzoli1998feed}, support vector machines \cite{mangasarian2000breast, van2007support}, etc., are used.

\begin{figure*}[!hb]
\centering
\includegraphics[width=1.0\textwidth]{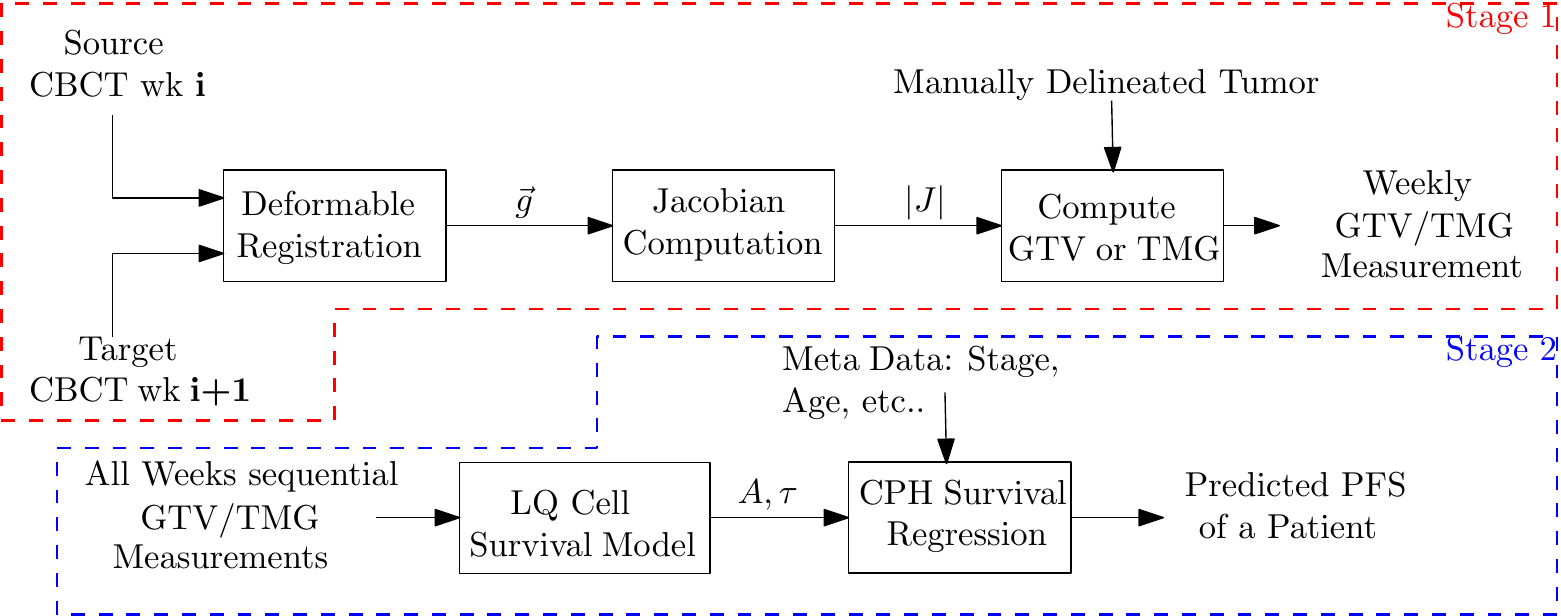}
\caption[Proposed Survival Regression approach]{A block diagram showing an overview of the proposed method for survival regression.}
\label{fig:Method_Block_Diagram}
\end{figure*}

Survival prediction was performed using features obtained from CT images \cite{hawkins2014predicting}.
Texture, geometry, and intensity based features were used in this work for classification with decision tree and support vector machines.
Here, instead of regression, patients were classified into long and short survival classes.
Long and short survival classification \cite{Chaddad2017} was also reported, using random forests.
Similar approaches have been reported \cite{DeJong2018}, where it was found that the predictability was better in stage I-III patients than in stage IV patients.
In another approach using CT images \cite{Zhang2017}, several classifiers along with feature selection methods were compared to evaluate the best combination.
It was found that principal component analysis for feature selection along with random forests gave the best predictive performance.
Other imaging modalities like Positron Emission Tomography (PET) were used \cite{Bissonnette2018}.
Here, serial 4D data of PET/CT were used and the features collected were the GTV and standard uptake value (SUV) along the several weeks of treatment.
Using statistical tests like Student t-test, Mann-Whitney U-test, and the log-rank test of statistical significance, the effect of features on the predictive performance was observed.
Similarly, PET images were used for predicting PFS \cite{Horne2014}. It was observed that the maximum SUV was a significant marker for reliable prediction.
CBCT images were used along with manually delineated GTV for survival analysis \cite{Jabbour2015} using the Cox's Proportional Hazard (CPH) model \cite{cox2018analysis}.
On 38 patients, it was reported that the risk of death decreased by 44.3\% for every 10\% decrease in the GTV.
CBCT images were also used for predicting the GTV at end of treatment using tumor regression on the GTV at one-third and two-thirds of the treatment period \cite{Brink2014}.
The survival curves showed significant differences between smaller and larger regression of the GTV.
GTV delineated in CT was propagated to CBCT \cite{van2017survival} and features were extracted from CT and CBCT images.
Features like tumor intensity, texture, wavelet, Laplacian of Gaussian and shape were extracted from the voxels within the GTV of both CT and CBCT.
A linear predictor \cite{aerts2014decoding} was obtained between the features in CT and CBCT for separating patients with good prognosis and bad prognosis.
Cox's regression was performed on this linear predicted value as a feature and a good concordance index of their approach was observed.

% Tumor growth
The linear quadratic (LQ) model \cite{fischer1971mathematical} was one of the earliest mathematical models to capture changes in tumor cells during radiotherapy.
The interaction of radiation with a quantity of cell population, resulting in cell multiplication and decay are studied \cite{chadwick1973molecular}, \cite{fowler1989linear}.
The LQ model can be used to express the survival fraction of tumor cells in a simple analytical form and in the prediction of survivability of a patient expressed as a tumor control probability (TCP).
TCP is expressed in terms of the survival function, cell death, and multiplication parameters \cite{zaider2000tumour} while making use of the LQ model.
The TCP is modified \cite{ORourke2009} by taking the repopulation of tumor cells into account.
Bayesian techniques were employed for adaptive prediction of tumor volumes \cite{Tariq2016} with radiation treatment.
Two cell population models were developed \cite{tariq2015mathematical}, to predict GTV changes.
A different mathematical model was proposed \cite{Stein2008}, where a regression-growth equation was formulated for capturing the tumor dynamics with treatment.
Their analysis indicated that the survival of a patient correlated with the log of growth rate but not with the log of regression/decay rate.
Using images, the volume of a tumor has been used to capture the tumor dynamics from a cell survival model \cite{wang2013mathematical}.
The effects of radiation treatment in terms of tumor growth and decay are specified by the cell survival model.
The least squares fit with the estimated and actual GTV was performed to obtain the tumor regression parameters.
These parameters were found to be in a finite range for different kinds of tumors.
Hybrid models \cite{Altrock2015} are also discussed in the literature.

Many of the works in the literature consider the GTV as a collection of homogeneous volume elements.
In the proposed method, the heterogeneous nature of each volume element is considered for more reliable analysis.

\subsection{Overview of the Proposed Model}
The goal of this work is to perform survival regression to predict the Progression Free Survival (PFS) of patients suffering from lung cancer.
The data for each patient consists of spatio-temporal sequences of CBCT images, with metadata like age (year), gender (Male/Female), stage of cancer (2A/2B/3A/3B), type of cancer (Adeno/Squamous Carcinoma), PFS (weeks), and the observation of patient death (0/1).
The median number of images in the spatio-temporal sequence is 7 corresponding to 6 weeks of treatment, where the first image was captured at the start of radiotherapeutic treatment.
Within the sequence of CBCT images, the data consists of manually delineated Gross Tumor Volumes (GTV), and Computed Tomography Number (CTN) at each voxel in an image.

Fig.~\ref{fig:Method_Block_Diagram} shows the sequence of steps that are followed to perform survival regression in the proposed method.
Initially, the images are registered to obtain deformation fields from which the Jacobian is computed.
GTV delineated by experienced radiologists was used here, rather than using auto-segmented GTV. CTN values are available from the image data as intensity values in an image.
A model is proposed to combine the information from GTV and CTN to produce a new reliable feature called Tumor Mass in GTV (TMG). %an Aggregation of Jacobian weighted CTN over the tumor GTV (JCTN). %Average Change of Mass per unit Volume (ACMV). %Jacobian corrected CTN (JCTN).
The sequence of images produces a set of TMG values which vary for every patient. The processing in stage 1 is shown in Fig.~\ref{fig:Method_Block_Diagram}.
In the past, Linear-Quadratic (LQ) cell survival model \cite{chadwick1973molecular} was used to estimate the GTV based on the initial GTV.
Instead, we use the TMG to obtain more reliable measurements when used along with the cell survival model.
The LQ cell survival model is also briefly discussed.
The parameters obtained from the least squares fit is used as a representative data point corresponding to a patient.
In the second stage, these parameters are used as features along with the metadata like age, gender, stage of cancer, and type of cancer, to predict the PFS of a patient.
For this, Cox's Proportional Hazard (CPH) survival regression model \cite{cox2018analysis} is used.

\section{Method}
\subsection{Modeling of Tumor Mass}
The density of an inhomogeneous substance at any point $\vec{z}; \forall \vec{z} \in \Omega$ can be defined as the rate of change of mass with respect to volume.
Here, $\Omega$ is the support space of a substance.
Radiodensity can be similarly expressed in terms of a mass and volume of any substance at a voxel.
We wish to model the mass of a tissue/substance at a voxel in terms of density and volume element.
The relationship can be expressed as follows:
\begin{gather}
\begin{aligned}
\rho(\vec{z},t) = \emph{density} &= \text{rate of change of \emph{mass} w.r.t \emph{volume}} \\
&= \frac{dm(\vec{z},t)}{dv(\vec{z},t)}
\end{aligned} \nonumber
\end{gather}
Therefore, the mass of an infinitesimally small volume element $dv(\vec{z},t)$ denoted as $dm(\vec{z},t)$, at a location $\vec{z}$ in an image at time $t$ of the treatment can be further expressed mathematically as follows:
\begin{equation} 
  \begin{aligned}
    dm(\vec{z},t) = \rho(\vec{z},t) dv(\vec{z},t)
  \end{aligned}
\end{equation}
Here, $\rho(\vec{z},t)$ denotes the radiodensity at a location $\vec{z}$ in an image at time $t$.
Discretizing the above equation using initial mass $m_i$ and final mass $m_f$, and similarly for volume, the above equation reduces to:
\begin{equation} 
  \begin{aligned}
  \label{eq:MassEqn}
    m_f - m_i = \rho (v_f - v_i)
  \end{aligned}
\end{equation}
Here the parameters $\vec{z}$ and $t$ are ignored for simplicity.
The location $\vec{z}$ for a three dimensional image is in the form of three components $\vec{z} = (z_1,z_2,z_3)$, where $z_1$, $z_2$, and $z_3$ are locations of the 3D voxel in the image.
When a patient undergoes treatment there is a change in radiodensity, mass, and volume of underlying tissues visible in a sequence of CBCT images.

\subsection{Tumor Mass in GTV and Jacobian}
The radiodensity at a location in a CBCT image is the Hounsfield Unit (HU) intensity. It is also called the CT Number.
It is proportional to the relative linear attenuation coefficient of the material with respect to that of water.
The tumor mass at any location in a GTV region undergoes changes when treated by either chemotherapy or radiotherapy. 
There are two factors to be considered during the treatment:
\begin{enumerate}
 \item Radiodensity changes: The radiodensity at any location may either increase or decrease due to treatment \cite{Guckenberger2013,liu2014mathematical}.
An increase in radiodensity indicates that the underlying tumor is quiescent or becoming more aggressive.
A decrease in the radiodensity indicates that the underlying tumor cells are being damaged due to treatment. 
 \item Volumetric changes: The tumor also undergoes volumetric change, as the number of tumor cells are multiplying or killed.
Cell damages are excreted out of the system by the biological processes in the human body.
The tumor receives nutrients and oxygen from the human body to multiply itself.
Due to the inherent nature of biological processes, the geometry of the solid tumor is approximately spherical \cite{weiswald2015spherical}.
The spherical shape results in different locations in the tumor receiving different amounts of oxygen and nutrition. 
Hence, the cell multiplication process results in different quantities at different locations.
\end{enumerate}
Therefore, to capture the dynamics of the tumor mass, the dynamics of volume change $dv(\vec{z},t)$ at any location must also be considered.
Image registration is performed between consecutive weeks to compute the volume change using Jacobian.

The image registration process consists of rigid registration of a pair of CBCT images in the first stage, followed by their deformable image registration in the next stage.
The rigid registration is implemented using the built in algorithm in the Insight ToolKit (ITK) \cite{yoo2002engineering_itk}, which uses the concept of mutual information \cite{mattes2003pet} to find the optimal alignment.
A 3-level multiresolution scheme is employed here to first register the images at $\frac{1}{4}^{th}$ the original resolution.
The registration parameters obtained at this resolution are used to initialize the registration parameters at $\frac{1}{2}$ the original resolution.
Finally, using them the registration parameters at the full resolution are obtained.
The registration parameters consists of 3 translation and 3 rotation parameters with respect to the center of the image.
As a preprocessing step, the images were smoothed using gaussian smoothing with a standard deviation of 2.0, 1.0 at the resolutions of $\frac{1}{4}$, $\frac{1}{2}$ of the original resolution.
The probability density functions of the pair of CBCT images were computed using Parzen window technique, with a bin size of 50.
We used gradient descent optimization scheme to find the optimal parameters with the step size of 1.0, limiting to a maximum of 100 iterations.

The second stage of image registration involves deformable image registration.
We have used the a non-linear registration technique, where the symmetric Local Correlation Coefficient (LCC) \cite{Lorenzi2013} is used as the similarity measure. 
The output of the first stage of registration is used as the transformed input image to register onto the next CBCT image.
Deformable image registration is performed between consecutive weeks to compute the volume change using Jacobian.
A Jacobian determinant expresses itself as a ratio of initial and final volumes. 
Determinant of a Jacobian matrix $J(\vec{z})$, commonly called simple Jacobian, gives local expansion or contraction at a location $\vec{z}$.
The Jacobian of a deformation field $\phi$ is defined as:
\begin{equation}
%\begin{align}
\label{eq:JacobianDet}
J(\vec{z}) = \begin{bmatrix} \frac{\partial \phi_1(\vec{z})}{\partial z_1} & \frac{\partial \phi_1(\vec{z})}{\partial z_2}
  & \frac{\partial \phi_1(\vec{z})}{\partial z_3} 
  & \\ \frac{\partial \phi_2(\vec{z})}{\partial z_1} & \frac{\partial \phi_2(\vec{z})}{\partial z_2} & \frac{\partial \phi_2(\vec{z})}{\partial z_3} 
  & \\ \frac{\partial \phi_3(\vec{z})}{\partial z_1} & \frac{\partial \phi_3(\vec{z})}{\partial z_2} & \frac{\partial \phi_3(\vec{z})}{\partial z_3}
\end{bmatrix}
%\end{align}
\end{equation}

Let us denote the determinant of a Jacobian matrix at time $t$ of treatment as $|J(\vec{z},t)|$.
Then,
\begin{gather}
  \begin{aligned}
    |J| = \frac{v_f}{v_i}
  \end{aligned} \nonumber
\end{gather}
The above expression, means that Jacobian is measuring the relative volume.
The initial volume of a tumor at any location in an image is the unit volume in the proposed approach.
However, while registering a spatio-temporal sequence of CBCT images, the volume of tumor undergoes deformation in the form of expansion or contraction.
After further manipulation, the above equation reduces to 
\begin{equation}
  \begin{aligned}
    |J| - 1 = \frac{v_f}{v_i} - 1 \\
    => |J| - 1 =  \frac{v_f - v_i}{v_i}
  \end{aligned} \nonumber
\end{equation}

\begin{equation}
  \begin{aligned}
  \label{eq:JacVol}
\therefore v_f - v_i = v_i (|J| - 1)
\end{aligned}
\end{equation}

Using Eq.~(\ref{eq:MassEqn}) and Eq.~(\ref{eq:JacVol}), the mass of substance at a volume element can be expressed as:
\begin{equation} 
  \begin{aligned}
    m_f = m_i + \rho_i v_i (|J| - 1)
  \end{aligned} \nonumber
\end{equation}

In the proposed approach, time $t$ takes on the values $0$, $1$, $\cdots$, $6$, corresponding to a pre-treatment CBCT image and six weeks of CBCT images taken during treatment.
Consecutive spatio-temporal image sequences of a patient are registered.
The implication of this approach is that in each registration, the initial volume of a voxel is $1.0$.
Also, the initial mass $m_i$ is therefore, $m_i = \rho_i v_i = \rho_i$.
Therefore the above equation reduces to:
\begin{equation} 
  \begin{aligned}
    m_f = \rho_i + \rho_i (|J| - 1)
  \end{aligned} \nonumber
\end{equation}
\begin{equation}
  \begin{aligned}
  \label{eq:final_mass_eqn}
\therefore m_f = \rho_i |J| 
\end{aligned}
\end{equation}

To compute the total tumor mass $M(t)$, a sum over all the voxels in the GTV is taken.
Let us define $\Omega_G$ as the support space where the tumor is active, which corresponds to locations within the GTV of the patient.
Using Eq.~(\ref{eq:final_mass_eqn}), the total mass of tumor within the GTV at time $t$ of the treatment can be re-expressed in discrete form as:
\begin{equation} 
  \begin{aligned}
  \label{eq:TMGVol}
    M(t) = \sum_{\forall \vec{z} \in \Omega_G} \rho_i(\vec{z},t) |J(\vec{z},t)|
  \end{aligned}
\end{equation}
The above expression computes the mass of tumor that is active while combining the information regarding the CT number as well as the volume change over the treatment period.
The tumor mass in GTV $M(t)$, abbreviated as TMG, is obtained from the above expression.
A sequence of TMG values is obtained for each patient corresponding to the several weeks of treatment.

Due to the inherent nature of CBCT image acquisition, the radiodensity is prone to noisy measurements.
Hence, the use of radiodensity at all the locations within the GTV is also erroneous.
Therefore, the computed TMG is also prone to errors. 
Instead, using Eq.~(\ref{eq:TMGVol}) consider the average TMG ($M_{avg}$) defined as follows:
\begin{equation} 
  \begin{aligned}
  \label{eq:AvgTMGVol}
    M_{avg}(t) = \frac{1}{|\Omega_G|} \sum_{\forall \vec{z} \in \Omega_G} \rho_i(\vec{z},t) |J(\vec{z},t)|
  \end{aligned}
\end{equation}
The average TMG is a better measure as the noisy measurements in mass averages out.
The TMG/Average TMG values are fitted with a cell survival model using least square error (LSE) estimation to obtain parameters representing the nature of change in TMG.
These parameters are used as covariates for survival regression.

\subsection{Mathematical modeling of Tumor Changes}  \label{sec:cell_survival}
Researches in molecular theory of cell responses to irradiation observe that radiation leads to breakage of molecular bonds.
These bonds may further repair themselves with time. The critical molecule that leads to cell death is the DNA (Deoxyribonucleic acid).
Single strand breakage may not lead to cell death, however, double-stranded breakage causes cell death.
The single-stranded breakage is found to repair itself by copying, whereas no such behavior was observed for the double-stranded breakage. 
The linear quadratic model of cell survival with radiation was proposed in \cite{chadwick1973molecular}.
Here, the survival is proportional to a linear combination of the single and double-stranded breakages.
The goal of this model is to estimate the tumor volume dynamics with radiation dose.
We discuss the LQ model in brief here.

According to the linear-quadratic model in radiobiology, the average number of DNA double-strand breaks ($N$) by ionizing
radiation in a single cell is assumed to be a linear-quadratic \cite{wang2013mathematical} function of radiation dose per fraction($D$).
The surviving proportion of cells ($S$) in the survival curve for a fully oxygenated cell population can be expressed by the following:
\begin{equation}
\label{eq:linear_quadratic}
\begin{aligned}
S = e^{-(AN + B)} = e^{-A(\alpha D + \beta D^2) - B}
\end{aligned}
\end{equation}
where A accounts for DNA single strand break and cell proliferation before reproductive death.
B accounts for the impact caused by the events that are not included in $\alpha$ and $\beta$ events, such as undefined factors related to cancer staging, histological grade, oxygen supply, blood flow, and the living environment after irradiation.
Here, $\alpha$ and $\beta$, are the cell and radiation specific radiobiological constants pertaining to single strand break and double-strand break, respectively.
They are determined for different types of cancers from cancer biology studies.
From radiobiology studies of tissue interaction with radiation, the survival fraction is proportional to the number of cells alive.
The above model, suggests that survival is proportional to the dose at very low doses only. At higher doses, the quadratic term dominates.
This is necessary, as low doses are required for radio-biological protection of tissues which are neighbors of the tumor.

Let the initial number of cells before radiation treatment be denoted as $N_{0}$.
Then, the initial tumor volume before radiation treatment is $V_0 = \tau \times N_{0}$.
The volume of tumor is assumed to be proportional to the number of cells.
$\tau$ is the proportionality parameter that is estimated from the data.
The value of $\tau$ changes from patient to patient.
It depends on several factors as noted below:
\begin{itemize}
 \item The number of cell deaths per unit volume and the capacity of the tumor cells to regenerate vary among patients \cite{wang2013mathematical} depending on the amount of radiation, age, general health and immunity, lifestyle and several other factors.
 \item Depending on the size and tumor shape, the amount of oxygen and nutrition that solid tumor receives varies significantly \cite{wang2013mathematical}. These factors affect the cell to volume ratio, as the cells receiving more oxygen and nutrition are likely to regenerate faster.
\end{itemize}

Once radiation is delivered to a patient, the number of cells changes.
While a given volume of the tumor can contain alive and dead cells, the volume of the tumor can be approximated from the survival fractions.
The number of living cells and dead cells after the $i^{th}$ week of radiation treatment can be given as, 
\begin{equation}
\label{eq:living_dead_kthweek}
\begin{aligned}
N_{i}^{l} = N_{i-1}^{l} \times S \\
N_{i}^{d} = N_{i-1}^{d} + N_{i-1}^{l} \times (1 - S)
\end{aligned}
\end{equation}
The subscript $i$, denotes the state after the $i^{th}$ dose of radiation treatment, and the superscripts $l$ and $d$ denotes the living and the dead components, respectively.
Once radiation is given to the tumor, the total number of tumor cells in the tumor volume changes.

During a time interval ($\Delta t$) between the fraction of treatments, $i$ and $i + 1$, tumor cells are damaged, while some of the cells may repopulate.
The repopulation happens exponentially with a constant $\kappa$.
The potential doubling time of living cells is usually $5.5$ days for lung tumor, $T_{2/1} = \frac{\ln 2}{\kappa}$.
Similarly, the decay of damaged cells also happens exponentially with a decay constant $\nu$.
The potential volume halving time of living cells is given as, $T_{1/2} = \frac{\ln 2}{\nu}$.
By considering the potential doubling and halving times of repaired and damaged cells, respectively, the total tumor volume at the beginning of the $(i+1)^{th}$ fraction is,
\begin{equation}
\label{eq:totalcells}
\begin{aligned}
N_{i} = {N}_{i}^{l} e^{\kappa \Delta t_i} + {N}_{i}^{d} e^{-\nu \Delta t_i}
\end{aligned}
\end{equation}
Since the tumor volume is proportional to the number of cells in the tumor, we can write:
\begin{equation}
\label{eq:totalvolume}
\begin{aligned}
V_{i} = N_{i} \times \tau
\end{aligned}
\end{equation}

The above mathematical model can be used to simulate the volume of tumor $V_{i}$ at time $i$, by assuming an initial volume $V_0$ obtained from the observed volume in a pre-treatment CBCT image. 
Estimated $V_{i}$ and actual observed volumes from the images can then be fitted by using least square error (LSE) estimation to obtain parameters that describe the dynamics of tumor volume changes.
The goodness of fit $R^2$ between measured and observed volumes is computed as follows:
\begin{equation}
\label{eq:goodnessfit}
\begin{aligned}
R^2 = 1 - \frac{ \sum{(y - \hat{y})^2} }{ \sum{(y - \bar{y}})^2 }
\end{aligned}
\end{equation}
where $y$ is a measured value, $\hat{y}$ is the predicted value, and $\bar{y}$ is the mean of measured data.

\subsubsection{Fitting the model}\label{model_fitting}
Solution to Eq.~\ref{eq:totalvolume}, gives the estimated volume at the end of each fraction.
We perform the least-squares technique to minimize the soft L1 loss and obtain the unknown parameters $A$, $B$, and $\tau$, that best explain/fit the actual volumes.
The residuals expressed as mean square error (MSE) between the observed and predicted values was minimized using the soft-L1 loss function.
\begin{equation}
\label{eq:mse}
\begin{aligned}
MSE = \frac{ \sum{(y - \hat{y})^2}} {W}
\end{aligned}
\end{equation}
Here, $W$ is the number of weeks a patient was treated.
The soft-L1 loss is expressed as: $2(\sqrt{1 + MSE} - 1)$.
The observed and predicted values in Eq.~\ref{eq:goodnessfit} can be either the GTV or the TMG.
The previously discussed cell survival biological model is fitted to these quantities.\\ \\
\emph{\textbf{Using GTV:}}\\
The data consists of GTV of $51$ patients who have undergone radiotherapy for a median period of $6$ weeks.
Each patient underwent a dosage of $10$\emph{Gy (Gray)} per fraction of one week duration.
Using the least-squares technique, the above discussed model was fit with the GTV for each patient.
Let us set the initial volume to $V_0$, the GTV in the CBCT before treatment.
Hereafter, the least square fit is tried with different values of the parameters under sensible constraints.
The constraints are that the value of $A$ and $B$ cannot be infinite. The value of $\tau$ is empirically limited to $[0.2, 2.0]$ which ensures finite solutions. 
Trust region method \cite{nocedal2006numerical} was used as the optimization scheme to search for the best fit.
A similar approach was used when using TMG as the quantity of interest.\\ \\
\emph{\textbf{Using TMG only:}}\\
Instead of considering the above model for GTV, the model for TMG is considered.
Here, it is considered that all source images had the unit volume as the initial volume at a voxel.
Therefore, in this scenario, Eq.~(\ref{eq:final_mass_eqn}) and Eq.~(\ref{eq:TMGVol}) hold true.

\section{Survival Regression}
The time till the event occurs is usually called the \emph{Survival time}. It is also called the \emph{waiting time}.
In some cases, the event may or may not happen. If the event does not happen, then that data is considered to be \emph{censored}.
Censoring can also happen due to the following:
\begin{itemize}
 \item The study ends.
 \item Discontinued followup.
 \item Withdrawal from study.
\end{itemize}
Censoring for the $i^{th}$ patient can be expressed using indicator variables as:
$\\ \\
\delta_i = \left\{
\begin{array}{ll}
1  & \mbox{if } \text{survival time} \leq \text{censoring time} \\
0 & \mbox{if } \text{survival time} \geq \text{censoring time}
\end{array}
\right. \\ \\
$
Here, $1$ represents uncensored data, while $0$ represents censored data, and censoring time is the time at the end of a study.

In survival regression, the goal is to regress different features/covariates against the period of survival.
We consider the death event in survival analysis to be the progression of a tumor.
Therefore, the time between the completion of treatment and the progression of a tumor is the survival time.
It is clear from the definition that this is equivalent to the PFS.
The proportional hazard model assumes that the hazard function for any patient is directly proportional to the baseline hazard.
The baseline hazard is independent of the covariates and dependent only upon time.

The Cox's Proportional Hazard (CPH) model \cite{cox2018analysis} aims to represent the hazard rate $\lambda(t|\vec{x}_i)$ as a function of time $t$ and covariates $\vec{x}_i$.
Let $\vec{x}_i$, $\forall i \in [1,N]$, denote the covariates of the $i^{th}$ patient with components $x_{i1}$, $x_{i2}$, $\ldots $ , $x_{iD}$, where $D$ is the dimensionality of the covariates.
The CPH model aims to represent the hazard rate $\lambda(t|\vec{x}_i)$ as a function of time $t$ and covariates $\vec{x}_i$.
It is parameterized by coefficients $b_j$, $\forall j \in [1,D]$, and is defined as:
\begin{equation} 
  \begin{aligned}
  \label{eq:CPH}
  \lambda(t|\vec{x}_i) = b_0(t) exp \Big( \sum_{j = 1}^{D} b_j x_{ij} \Big)
  \end{aligned}
\end{equation}
Here $b_0(t)$ is the baseline hazard.
The intuitive idea behind this model is that the log-hazard of a patient is a linear function of the features $\vec{x}_i$, and the baseline hazard.
The baseline hazard is a population-based quantity that changes only with time, while the exponential part is time-invariant and only dependent on $\vec{x}_i$.
The overall hazard is, therefore, a scaling of the baseline hazard, where the exponential part is the scale factor specific to each patient.

\begin{figure*}[ht]
\centering
\includegraphics[width=1.0\textwidth]{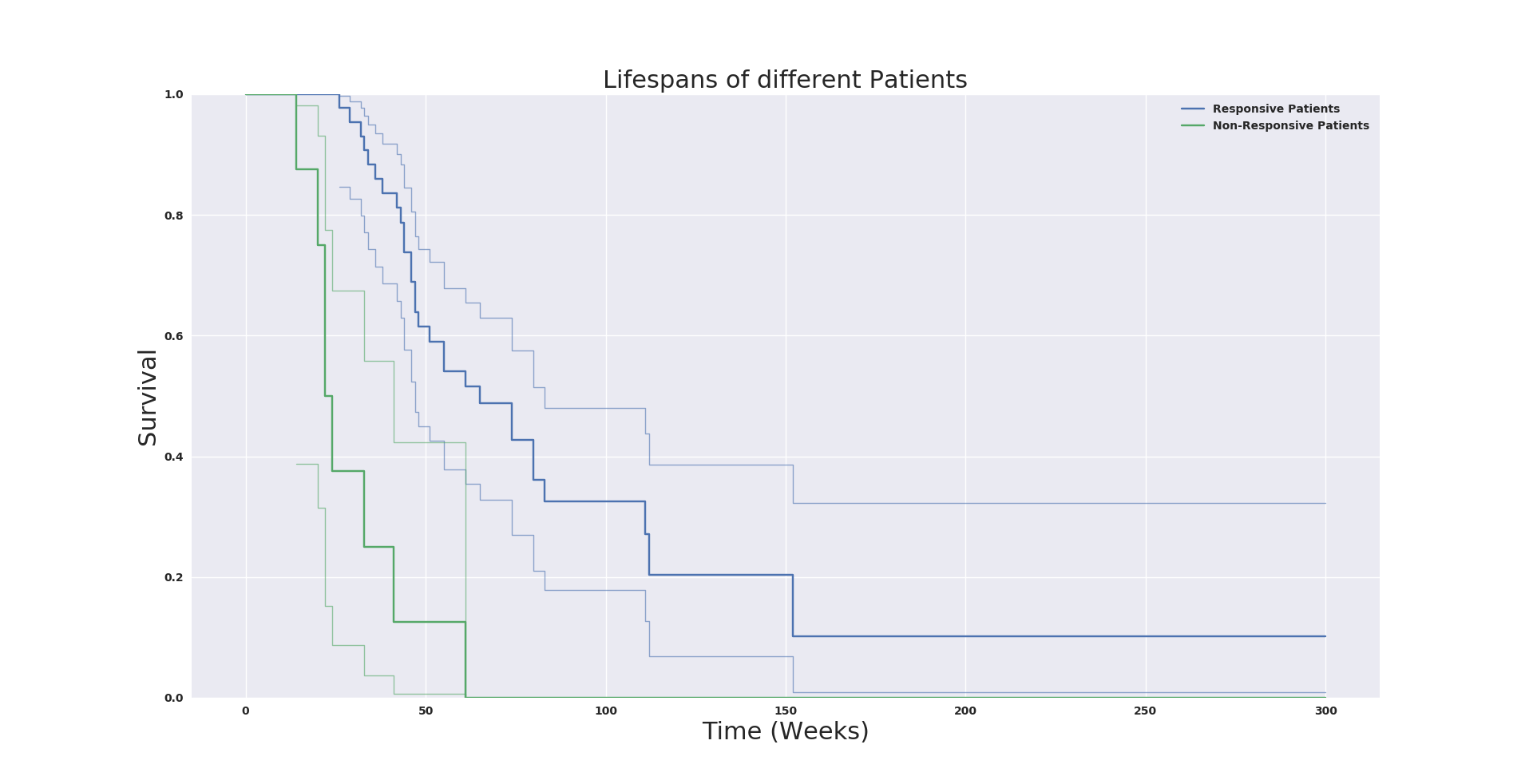}
\caption{Survival curve of responsive and non-responsive patients using Kaplan-Meier estimate. Figure shows the 95\% confidence intervals using thin lines.}
\label{fig:KMFCurve_RNR}
\end{figure*}

\begin{figure*}[t]
\centering
\includegraphics[width=1.0\textwidth]{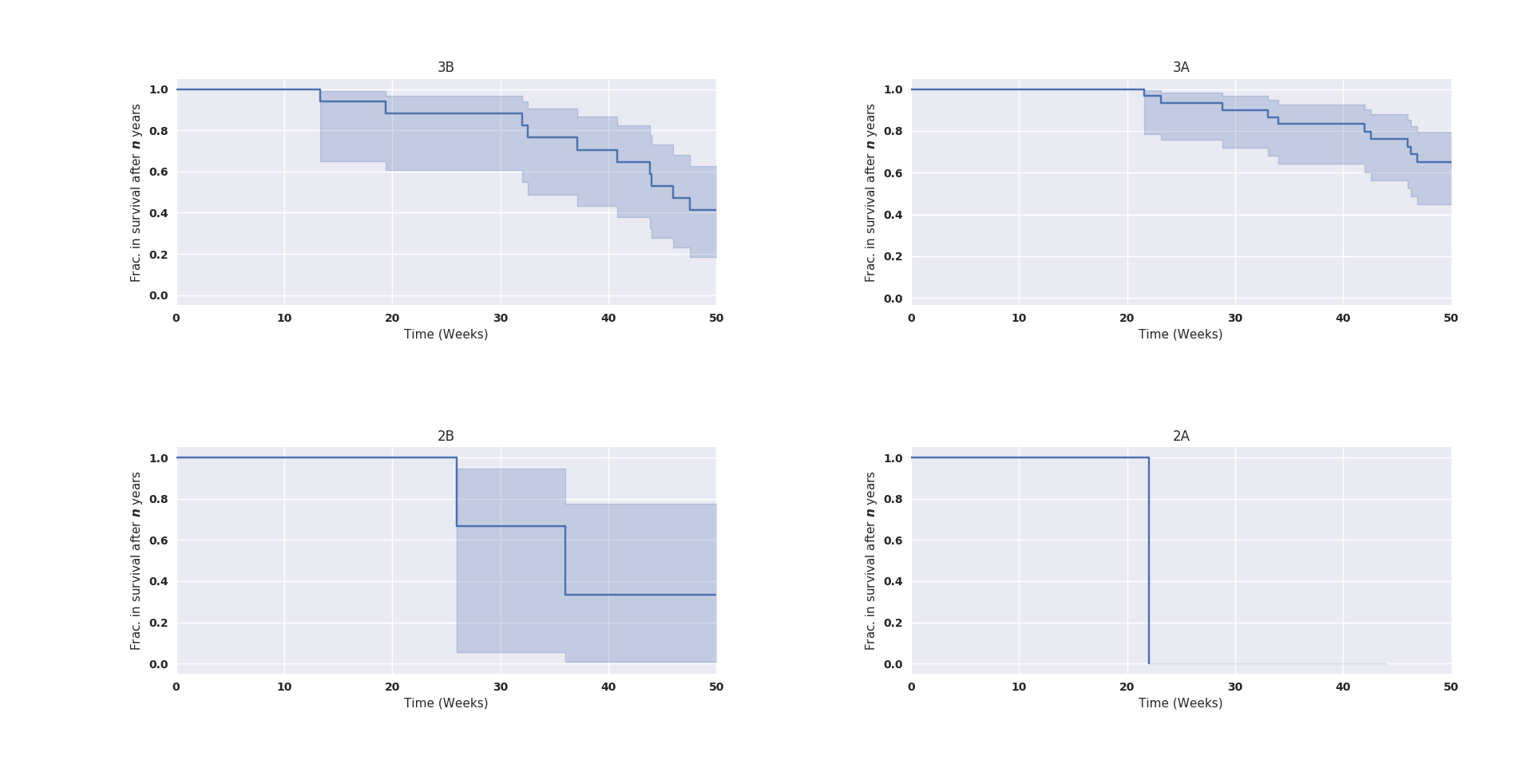}
\caption{Survival curves of patients in different stages of cancer.}
\label{fig:Survival_Stage}
\end{figure*}

Let $X$ be a non-negative random variable representing the lifetimes taken from the population under study. 
Let $d_t$ be the number of patients with progression of tumor at time $t$, and $R(t)$ be the set of patients at risk at time $t$ defined as $R(t) = \{ i : X_i \geq t \}$.
The partial likelihood function of a patient is the conditional probability that the $i^{th}$ patient has progressed given that this patient is one of the patients from the risk set $R(t)$.
It can be expressed as follows:
\begin{equation} 
  \begin{aligned}
  \label{eq:CPHLikelihood_Def}
  L_n(\vec{b}) = P(\text{$i^{th}$ patient progressed} | \text{one progression from }R(t))
  \end{aligned}
\end{equation}
The overall likelihood can be expressed as:
\begin{equation} 
  \begin{aligned}
  \label{eq:CPHLikelihood}
  L(\vec{b}) = \prod_{i : \delta_i = 1} \frac {exp( \vec{b}^T \vec{x}_i )} { \sum_{j : R(t)}  exp( \vec{b}^T \vec{x}_j ) }
  \end{aligned}
\end{equation}
The baseline hazard, $b_0(t)$ need not be estimated in CPH model as it cancels out in the likelihood.
However, the baseline hazard can be obtained by setting the covariates $\vec{x}_i$ to zero.
The baseline hazard at time $t$ can also be estimated using Breslow's method \cite{breslow1972discussion} as:
\begin{equation} 
  \begin{aligned}
  \label{eq:Baseline_Hazard}
  b_0(t) = \frac{d_t} { \sum_{j : R(t)}  exp( \vec{\hat{b}}^T \vec{x}_j ) }
  \end{aligned}
\end{equation}
Here, $\vec{\hat{b}}$ is the maximum likelihood estimated coefficients, obtained from the partial likelihood function expressed in Eq.~\ref{eq:CPHLikelihood}.
The solution of the above model involves, determining the coefficients $b_j$, $\forall j \in [1,D]$, where $D$ is the number of features.
The parameters $\vec{\hat{b}}$ are estimated by maximizing the partial likelihood.

In this method, the parameters obtained from the cell survival model discussed in the previous section ($A$, $B$, $\tau$) are used here as the features $\vec{x}_i$.
Additionally, the use of other information about a patient, such as age, gender, stage and type of cancer are also considered in regression.

\begin{figure*}[!htp]
\centering
\includegraphics[width=1.0\textwidth]{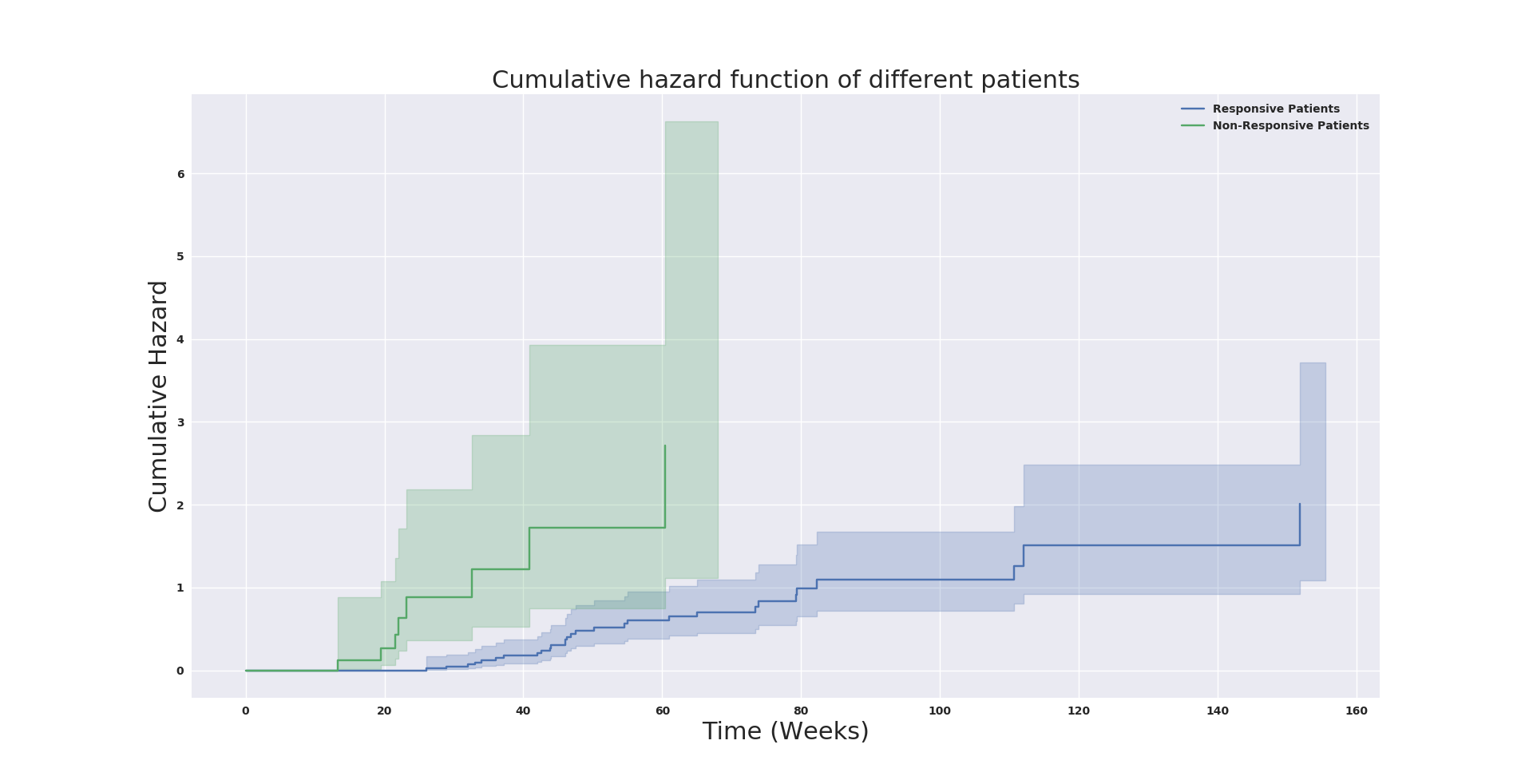}
\caption{Cumulative Hazard curve of responsive and non-responsive patients using Nelson-Aalen estimator. Figure shows the 95\% confidence intervals using thin lines.}
\label{fig:Hazard_NAF_RNR}
\end{figure*}

\section{Results and Discussion}
\subsection{Dataset}
The datasets for this study were obtained from a larger cohort of 251 patients with lung cancer, who were treated using radical radiotherapy with
curative intent \cite{shrimali2018impact}. The survival outcomes for this group have been reported with subgroup analyses for sequential chemoradiation,
concurrent chemoradiation and radiotherapy alone \cite{shrimali2018impact}. Analyses have also been reported for patients treated using 3-dimensional conformal
radiotherapy (3D-CRT) versus volumetric modulated radiotherapy (VMAT) and patients with larger target volumes (PTV $> 500ml$) versus smaller
target volumes (PTV $< 500ml$) \cite{shrimali2018impact}. It has been shown that the treatment protocol \cite{shrimali2018impact} has resulted in outcomes that are at par
with published literature from large multi-centre phase-III clinical trials (RTOG 0617) as well as retrospective data from the UK \cite{iqbal2019hypofractionated, bradley2015standard, chun2017impact, arunsingh2018survival}.
From this big dataset of radically non small cell lung cancer (NSCLC) patients treated in a tertiary care center, fifty-one random patients who received conventionally fractionated RT over 6 weeks were selected. All of these patients had completed their routine treatment using concurrent chemoradiation (60Gy in 30 fractions over six weeks, with concurrent cisplatin and etoposide). Patients treated with hyperfractionated accelerated RT and hypofractionated accelerated RT were excluded from this study in order to achieve uniformity in the selected group and allow us to test the three week prediction model as proposed above.
The Tata Medical Center (Cancer Hospital) - Institutional Review Board discussed our study using CBCT from archived data of patients and did not recommend informed consent from patients for this study.
All protocols and methods in the study were in agreement with the guidelines and regulations.

Our dataset consists of 51 patients, among whom 29 patients had stage IIIA, 18 had stage IIIB, 1 patient had stage IIA and 3 patients had stage IIB cancer.
45 patients were males and 6 patients were females with a median age of 62 years (range, 33-81 years).
CBCT images were acquired on the Varian Medical Systems OBI CBCT device using low-dose thorax settings described as 80kV, 25mA, pulse width of 8 milliseconds, small focal spot, on a half fan setting.
Standard Ram-Lak filter was applied to the sinogram prior to CBCT image reconstruction using the FDK algorithm \cite{feldkamp1984practical}.
All patients had a follow-up contrast enhanced CT scan done 3 months post therapy completion and Radiology Response Evaluation Criteria in Solid Tumors (RECIST) \cite{eisenhauer2009new} criteria were used as reference for therapy response.
In our dataset, the birth event is the completion of radiation treatment, and the death event is the progression of a tumor in that patient.
Censoring occurs in this case when the patient has had no progression of a tumor.
The progression of a tumor is observed by the radiation oncologists, during the follow-up CT scan, where the RECIST criteria are also determined based on the size of the tumor.

\subsection{Image Registration}

\begin{figure*}[!ht]
\centering
\includegraphics[width=0.5\textwidth,height=0.5\textwidth]{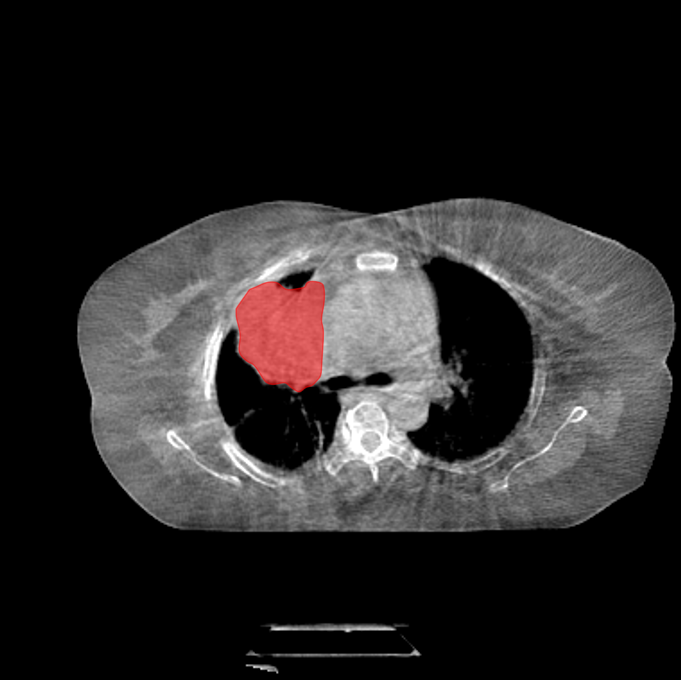}
\caption{A sample slice with GTV annotated in red is shown.}
\label{fig:CBCTAnnotated}
\end{figure*}

\begin{figure*}[!ht]
\centering
\subfloat[]{\includegraphics[width=0.33\textwidth,height=0.33\textwidth]{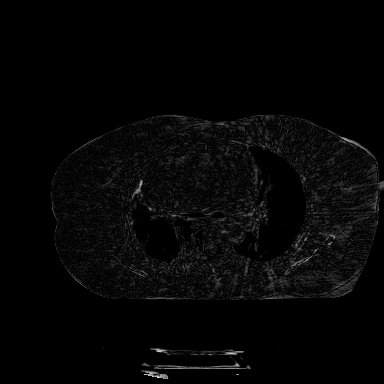}
\label{fig:CA}}
\subfloat[]{\includegraphics[width=0.33\textwidth,height=0.33\textwidth]{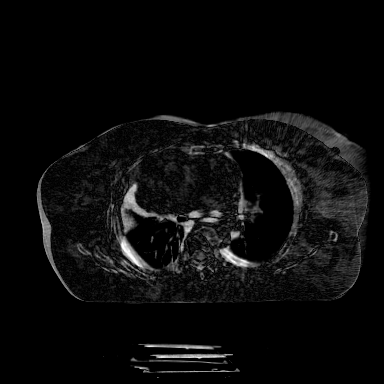}
\label{fig:CU}}
\subfloat[]{\includegraphics[width=0.33\textwidth,height=0.33\textwidth]{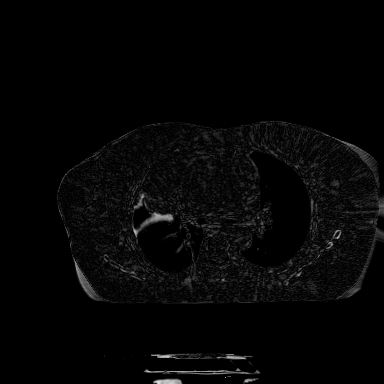}
\label{fig:BL}}
\caption{Difference images of a pair of registered images of a patient are shown. Slices consist of (a) Clinically acceptable, (b) Clinically unacceptable, and (c) Borderline registrations.}
\label{fig:Reg_Quality}
\end{figure*}

The registered images were manually checked for any misalignments by experienced radiation oncologists.
The image registration technique used in this study was evaluated objectively. Four different anatomical structures were taken into consideration while evaluating the registered images. All the registered images of each patient were assessed for the accuracy of registration. A scoring system was used for the same.
The anatomical structures selected were vertebrae, ribs, sternum and gross tumour volume. The CBCT image from the first week was used to determine the location of the GTV in each patient. The difference images which were created by the registration system were used for evaluation of the accuracy of registration.
\begin{enumerate}
 \item Vertebrae fusion scores: 
 \begin{itemize}
  \item 1 : vertebrae showed perfect fusion
  \item 0 : poor fusion
 \end{itemize}
\item Ribs fusion scores: 
\begin{itemize}
  \item 2 : ribs showed perfect fusion
  \item 1 : near perfect fusion
  \item 0 : bad fusion
\end{itemize}
\item Sternum fusion scores: 
\begin{itemize}
  \item 2 : sternum showed perfect fusion
  \item 1 : near perfect fusion
  \item 0 : bad fusion
\end{itemize}
\item Gross tumour volume scores:
\begin{itemize}
  \item 3 : Difference image showed perfect match.
  \item 2 : Rim of hyperdensity at the area around GTV in the difference image
  \item 1 : Large area of hyperdensity near the GTV area in the difference image
\end{itemize}
\end{enumerate}
The scores were added and each registered image was given a particular score. The criteria for accepting a registered image clinically was determined as 
\begin{itemize}
  \item Clinically unacceptable : 1-3
  \item Borderline : 4-5
  \item Clinically Acceptable : 6-8
\end{itemize}
Above ranges were decided in consultation with clinical oncologists participated in this study.
A sample slice of a patient with the GTV delineated is shown in Fig.~\ref{fig:CBCTAnnotated}.
Samples slices of a patient after registration are shown in Fig.~\ref{fig:Reg_Quality}.
In this patient, the registration quality for different weeks was found to have all the three above categories.
As seen in Fig.~\ref{fig:CA}, the vertebrae, ribs, sternum and GTV have registered properly with minimal error of less than 1 voxel width.
In Fig.~\ref{fig:CU}, there is visible error in the registrations of vertebrae, sternum, ribs and GTV.
In the borderline case of Fig.~\ref{fig:BL}, the vertebrae, ribs and sternum have registered properly, while there is visible error in registration around the GTV.
It was observed that when the change in the GTV was large between the CBCT images taken between the two weeks the scores were low due to imperfect registration around the GTV areas.
It was also observed that when the field of view between the two image sets were different the registration was less perfect.
In our dataset, the registration was clinically acceptable in 85.81\% of the cases, border line in 10.14\%, and unacceptable in 4.05\% cases, respectively.

\subsection{Survival Analysis}
Survival analysis and regression were performed using the \emph{lifelines} package in \emph{python} programming environment.
Survival analysis using \emph{Kaplan-Meier Estimate} \cite{kaplan1958nonparametric} is applied to the dataset.
Median survival is defined as the time at which the survival function equals $0.5$ in the survival curve.
The median survival was found to be $54.57$ weeks.
Based on the RECIST criteria, patients can be grouped into responsive and non-responsive categories.
Responsive patients have RECIST as either Complete Response (CR) or Partial Response (PR), while non-responsive patients have RECIST as Stable Disease (SD) or Progressive Disease (PD).
It was found that the median survival in non-responsive patients was $24.0$ weeks,
while that of responsive patients it was $65.0$ weeks.
The survival curves for responsive and non-responsive patients is shown in Fig.~\ref{fig:KMFCurve_RNR}.
It can be observed from the figure that, the probability of survival in non-responsive patients is lower compared to that of the responsive patients.
Right from the completion of treatment, there was a high risk in non-responsive patients.
Both the curves seem to differ in the initial stages to a great extent.
The proportional hazards assume that the survival curves differ only by a constant multiplicative factor.
This assumption holds true for our dataset as the visual interpretation shows that the curves do not intersect differing only by a constant factor.

The log-rank test was performed as in \cite{Wang2017} to compare the survival distributions of the responsive and non-responsive samples of patients with one degree of freedom.
The null hypothesis is that there is no difference between the two survival curves.
The distribution of the log-rank statistic is considered to follow the Chi-Squared distribution.
The test statistic was 28.11 with a p-value of 0.0.
Since the level of significance is less than a cut off value of 0.05 corresponding to 95\% confidence interval, we can reject the null hypothesis that both the survivals are from the same distribution.

The survival curves are compared for different stages of cancer in the dataset.
Fig.~\ref{fig:Survival_Stage} shows the survival curves for 4 different stages of cancer patients in our dataset.
It can be observed from the curves that, among the higher stages of cancer (3A or 3B), the survivability of 3A is better than 3B.

Estimation of the cumulative hazard function is done using the \emph{Nelson-Aalen estimator} \cite{nelson1969hazard}.
For responsive and non-responsive patients the cumulative hazard curve is shown in Fig.~\ref{fig:Hazard_NAF_RNR}.
Clearly, the hazard curve of non-responsive patients is higher than responsive patients, suggesting that non-responsive patients are at a higher risk at any time.

\begin{figure*}[h]
\centering
\includegraphics[width=1.0\textwidth]{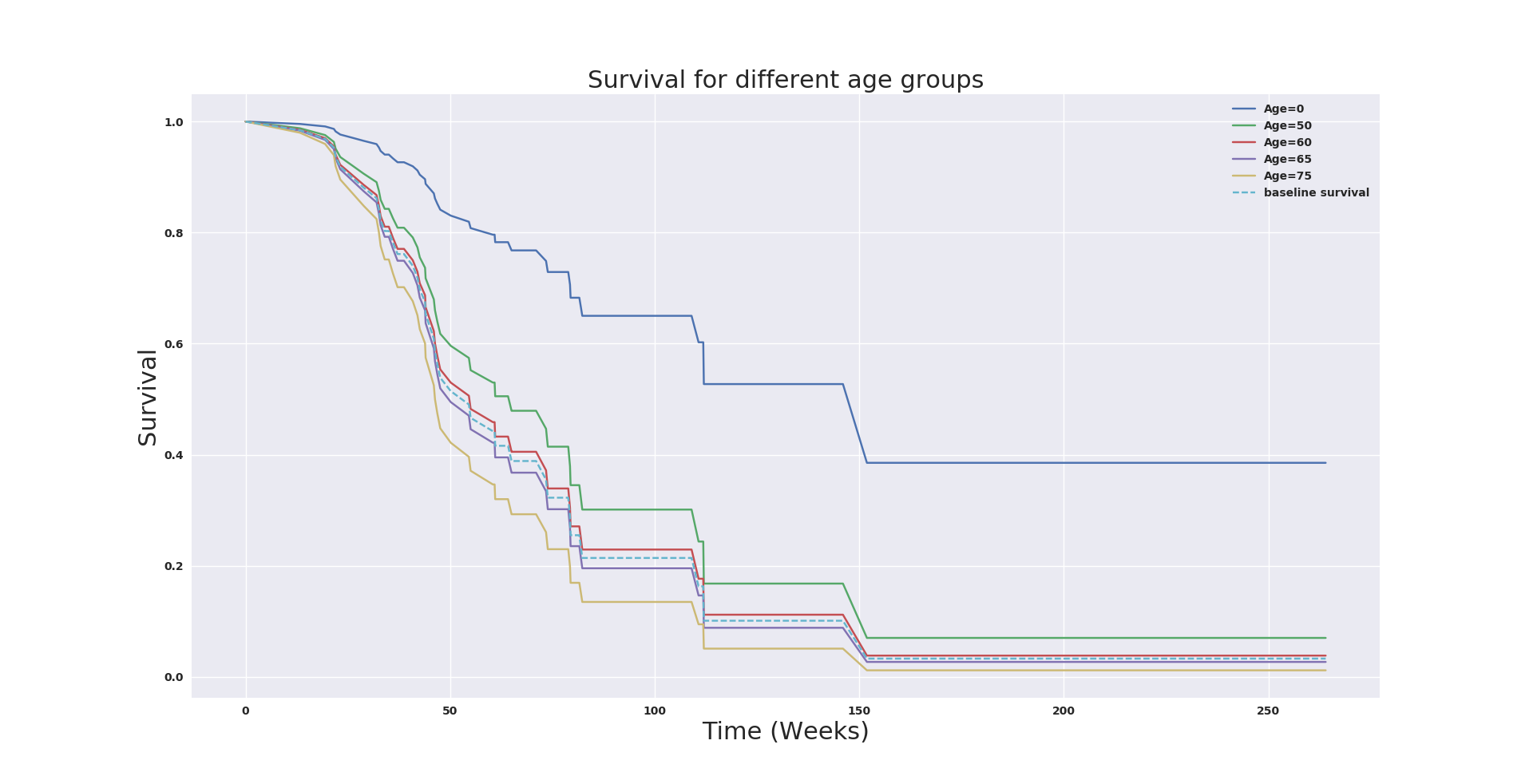}
\caption{Survival curves of different age groups when compared to the baseline survival.}
\label{fig:Survival_AgeGroups}
\end{figure*}

The survival of different groups of patients is compared based on their ages.
The survival curves for the 5 different age groups with baseline survival are shown in Fig.~\ref{fig:Survival_AgeGroups}.
It can be clearly seen that the survivability of patients is much better at a younger age compared to older patients.

\subsection{Cell Survival model}
The hyper-parameters \cite{wang2013mathematical} $\alpha$, $\beta$, $\kappa$, $\nu$ in the cell survival model, were set empirically.
Few values of these hyper-parameters are fixed based on the type of cancer and other governing factors \cite{ORourke2009, Guckenberger2013}.
According to the work reported in \cite{Guckenberger2013}, the hyper-parameter values are set as follows: $\kappa$ was chosen to be 5.5/7 $(Week)^{-1}$ and $\nu$ was $\kappa/3$.
$\alpha$ was 0.3 and $\beta$ was $\alpha/10$.
The same hyper-parameters $\alpha$, $\beta$, $\kappa$, $\nu$ in the cell survival model, were set while using GTV, average CTN, and average TMG.
The parameters estimated from the cell survival model are $A$, $B$ and $\tau$ \cite{wang2013mathematical}.

For both the above approaches using the GTV alone or the TMG alone, these parameters were kept constant and equal.
Only the parameters $A$, $B$ and $\tau$ and were found by minimizing the sum of squared error between the estimated and actual quantities as in Eq.~\ref{eq:mse}.
A threshold of $0.8$ is considered for the goodness of fit $R^2$.
It was observed that $41$ out of the $51$ patients have a good fitness in the case of GTV and $11$ good fits in the case of average TMG.
Even though there are fewer good fits when using average TMG, the survival prediction is better compared to using GTV or average CTN.

\subsection{Survival Regression}
Finally, in survival regression, the different features were $A$, $B$, $\tau$, age, gender, stage of cancer and type of cancer which correspond to features in the vector $x$.
One hot encoding was used to convert the categorical features to numerical features.
The Cox' Proportional Hazard model was used on all the resultant features for survival regression.
Newton Raphson's method was used as the optimization method with a step size of 0.25.
The objective function was the soft-L1 loss as discussed before.

Table.~\ref{tab:cph_summary} summarizes the results of survival regression.
Regression was performed with and without the meta information of patients.
The degrees of freedom are indicated in the table. 
Comparison of GTV, average CTN, TMG and average TMG are provided here.

\begin{table*}[]
\centering
\caption{Comparison of CPH model regression on using GTV, average CTN, TMG and average TMG with cell survival model.}
\label{tab:cph_summary}
\resizebox{1.0\columnwidth}{!}{%
\begin{tabular}{|c|c|c|c|c|c|c|c|c|c|c|c|c|}
\hline
\multicolumn{1}{|l|}{}                                             & \multicolumn{4}{c|}{{\color[HTML]{3166FF} GTV}}                                                                                           & \multicolumn{4}{c|}{{\color[HTML]{3166FF} Average CTN}}                                                                                   & \multicolumn{4}{c|}{{\color[HTML]{3166FF} Average TMG}}                                                                                                        \\ \hline
{\color[HTML]{3166FF} Covariates}                                  & {\color[HTML]{3166FF} c-index} & {\color[HTML]{3166FF} likelihood ratio} & {\color[HTML]{3166FF} p-value} & {\color[HTML]{3166FF} BIC}    & {\color[HTML]{3166FF} c-index} & {\color[HTML]{3166FF} likelihood ratio} & {\color[HTML]{3166FF} p-value} & {\color[HTML]{3166FF} BIC}    & {\color[HTML]{3166FF} c-index}       & {\color[HTML]{3166FF} likelihood ratio} & {\color[HTML]{3166FF} p-value}       & {\color[HTML]{3166FF} BIC}             \\ \hline
{\color[HTML]{3166FF} A,$\tau$ and meta-info (dof = 13)}           & 0.47                           & {\color[HTML]{000000} 14.53}            & {\color[HTML]{000000} 0.34}    & {\color[HTML]{000000} 275.24} & {\color[HTML]{000000} 0.54}    & {\color[HTML]{000000} 18.32}            & {\color[HTML]{000000} 0.15}    & {\color[HTML]{000000} 271.46} & {\color[HTML]{000000} \textbf{0.64}} & {\color[HTML]{000000} \textbf{20.2}}    & {\color[HTML]{000000} \textbf{0.09}} & {\color[HTML]{000000} \textbf{269.52}} \\ \hline
{\color[HTML]{3166FF} A,$\tau$ and without meta-info (dof = 2)}    & 0.56                           & {\color[HTML]{000000} 0.58}             & {\color[HTML]{000000} 0.75}    & {\color[HTML]{000000} 245.94} & {\color[HTML]{000000} 0.50}    & {\color[HTML]{000000} 0.74}             & {\color[HTML]{000000} 0.69}    & {\color[HTML]{000000} 245.79} & {\color[HTML]{000000} 0.52}          & {\color[HTML]{000000} 1.05}             & {\color[HTML]{000000} 0.59}          & {\color[HTML]{000000} 245.48}          \\ \hline
{\color[HTML]{3166FF} A,$\tau$, B and meta-info (dof = 14)}        & 0.46                           & {\color[HTML]{000000} 5.12}             & {\color[HTML]{000000} 0.98}    & {\color[HTML]{000000} 288.59} & {\color[HTML]{000000} 0.42}    & {\color[HTML]{000000} 11.51}            & {\color[HTML]{000000} 0.64}    & {\color[HTML]{000000} 282.20} & {\color[HTML]{000000} 0.46}          & {\color[HTML]{000000} 5.78}             & {\color[HTML]{000000} 0.97}          & {\color[HTML]{000000} 287.92}          \\ \hline
{\color[HTML]{3166FF} A,$\tau$, B and without meta-info (dof = 3)} & 0.58                           & {\color[HTML]{000000} 0.87}             & {\color[HTML]{000000} 0.83}    & {\color[HTML]{000000} 249.58} & {\color[HTML]{000000} 0.50}    & {\color[HTML]{000000} 0.60}             & {\color[HTML]{000000} 0.89}    & {\color[HTML]{000000} 249.85} & {\color[HTML]{000000} 0.54}          & {\color[HTML]{000000} 1.56}             & {\color[HTML]{000000} 0.67}          & {\color[HTML]{000000} 248.89}          \\ \hline
\end{tabular}}
\end{table*}

Bayesian Information Criteria (BIC) \cite{kass1995bayes} for censored data \cite{volinsky2000bayesian}, which is based on maximum partial likelihood estimation is used here for model selection.
A model with lower BIC is considered a better model. 
A difference of 5.72 is observed between using GTV and average TMG, which corresponds to a strong evidence \cite{kass1995bayes} that average TMG is a better measure.
Similarly, average TMG was found to be better than average CTN.
The likelihood ratio test (LR-Test) is performed to compare the regression model with and without the covariates.
The model without the covariates contained only the censoring information and the PFS.
The model with covariates contains the meta information of a patient and parameters obtained from the cell survival model.

Using the GTV, the LR is 14.53 in the best case with 13 degrees of freedom and a p-value of 0.34.
Using average CTN, the LR is 18.32 in the best case with 13 degrees of freedom and a p-value of 0.15.
It is found that using the TMG, the LR is 11.99 in the best case with 13 degrees of freedom and a p-value of 0.53.
However, when average TMG is used the LR is 20.2 in the best case with 13 degrees of freedom and a p-value of 0.09.

It is observed that the results are statistically significant when using average TMG, considering a significance level of 90\%.
$B$ is found to be an insignificant feature indicating that other external factors have no contribution within the cell survival model.
The c-index on the entire dataset using GTV with meta information is found to be 0.47 and is 0.54 when using CTN with meta information.
In the case of TMG with meta information, the c-index on the entire dataset is 0.58.
While using average TMG, the c-index improves to 0.64.

The tumor intensity variations though noisy in a CBCT image, suggest that there is heterogeneity in the tumor.
The texture of tumor compared to the surrounding soft tissues is slightly different when compared on different windows of the HU intensity scales.
Due to the inherent heterogeneity within a tumor, different volume elements have different intensities, and therefore reflect different activities at the cell microscopic level.
The GTV considers the total volume of the visible tumor but does not consider the contribution of each volume element.
The average TMG includes the effect of the radiodensity at each voxel and therefore takes into account the heterogeneity of the tumor.
Hence, from our experiments, the proposed model could better predict the PFS of a patient.

\section{Conclusion}
A simple model is proposed to incorporate the CTN number into the GTV, resulting in the computation of the mass of the tumor within the GTV (TMG) statistic.
The novel feature considers the image intensities and the Jacobian expansion at a voxel.
The Jacobian is obtained from the deformation field of the registered pair of images.
A cell survival model is used to obtain parameters describing the behavior of average TMG over a treatment period.
Survival analysis is performed to check for the proportional hazards assumption, which is found to hold on our dataset.
Survival regression results show that the average TMG is a better feature when compared using only the GTV.
The average TMG considers tumor heterogeneity into account while the GTV does not.
It is shown that average TMG can be used for prediction of PFS in lung cancer patients.

\section*{Disclosures}
Competing Interests: The authors declare that they have no competing interests.
All authors report grant from Ministry of Human Resources and Development (MHRD), Government of India, during the conduct of the study. 
There are no relevant financial activities in the submitted work. There have been no patents and there is no potential conflict of interest.

\section*{Acknowledgment}
This study was designed and performed by a collaboration between IIT Kharagpur and Tata Medical Center, Kolkata.
This study was carried out under a project (ref. NO. 4-23/2014-TS.I, Dt. 14-02-2014) sponsored by  Ministry of Human Resources and Development (MHRD), Government of India.
The funding sources had no role in the study design, data collection, analysis of interpretation, or the writing of this manuscript.

%%%%% References %%%%%

\bibliographystyle{unsrt}
\bibliography{references}

\end{document}